\documentclass[prl, twocolumn, superscriptaddress,
               preprintnumbers, amsmath, amssymb, amsfonts]{revtex4}
\usepackage{graphicx, dcolumn, bm, here}

\begin{document}

\preprint{UTCCS-P-31, UTHEP-542, KEK-CP-193, HUPD-0702, RBRC-666}

\title{Light quark masses from unquenched lattice QCD}

%%%%%%%%%%
\newcommand{\RBRC}{
  RIKEN BNL Research Center,
  Brookhaven National Laboratory,
  Upton, New York 11973,
  USA}

\newcommand{\Tsukuba}{
  Graduate School of Pure and Applied Sciences,
  University of Tsukuba, 
  Tsukuba 305-8571, 
  Japan}

\newcommand{\CCS}{
  Center for Computational Sciences, 
  University of Tsukuba, 
  Tsukuba 305-8577, 
  Japan}

\newcommand{\ICRR}{
  Institute for Cosmic Ray Research, 
  University of Tokyo, 
  Kashiwa 277-8582, 
  Japan}

\newcommand{\KEK}{
  High Energy Accelerator Research Organization (KEK), 
  Tsukuba 305-0801, 
  Japan}

\newcommand{\Sokendai}{
  School of High Energy Accelerator Science,
  The Graduate University for Advanced Studies (Sokendai),
  Tsukuba 305-0801,
  Japan}

\newcommand{\Hiroshima}{
  Department of Physics, 
  Hiroshima University,
  Higashi-Hiroshima 739-8526, Japan}

% authors

\author{T.~Ishikawa}
\affiliation{\CCS}
\affiliation{\RBRC}

\author{S.~Aoki}
\affiliation{\Tsukuba}
\affiliation{\RBRC}

\author{M.~Fukugita}
\affiliation{\ICRR}

\author{S.~Hashimoto}
\affiliation{\KEK}
\affiliation{\Sokendai}

\author{K-I.~Ishikawa}
\affiliation{\Hiroshima}

\author{N.~Ishizuka}
\affiliation{\CCS}
\affiliation{\Tsukuba}

\author{Y.~Iwasaki}
\affiliation{\Tsukuba}

\author{K.~Kanaya}
\affiliation{\Tsukuba}

\author{T.~Kaneko}
\affiliation{\KEK}
\affiliation{\Sokendai}

\author{Y.~Kuramashi}
\affiliation{\CCS}
\affiliation{\Tsukuba}

\author{M.~Okawa}
\affiliation{\Hiroshima}

\author{Y.~Taniguchi}
\affiliation{\CCS}
\affiliation{\Tsukuba}

\author{N.~Tsutsui}
\affiliation{\KEK}

\author{A.~Ukawa}
\affiliation{\CCS}
\affiliation{\Tsukuba}

\author{N.~Yamada}
\affiliation{\KEK}
\affiliation{\Sokendai}

\author{T.~Yoshi\'{e}}
\affiliation{\CCS}
\affiliation{\Tsukuba}

\collaboration{CP-PACS and JLQCD Collaborations}
\noaffiliation
%%%%%%%%%%

%\date{November 9, 2007}
\date{\today}% It is always \today, today,
             % but any date may be explicitly specified

\begin{abstract}
We calculate the light meson spectrum and the light quark masses by
lattice QCD simulation, treating all light quarks dynamically and
employing the Iwasaki gluon action and the 
nonperturbatively $O(a)$-improved Wilson quark action.
The calculations are made at the squared lattice spacings at an equal
distance $a^2\simeq 0.005$, $0.01$ and $0.015~{\rm fm}^2$,
and the continuum limit is taken assuming an $O(a^2)$ discretization error.
The light meson spectrum is consistent with experiment.
The up, down and strange quark masses in the
$\overline{\rm MS}$ scheme at $2~{\rm GeV}$ are
$\overline{m}=(m_{u}+m_{d})/2=3.55^{+0.65}_{-0.28}~{\rm MeV}$ and
$m_s=90.1^{+17.2}_{-6.1}~{\rm MeV}$ where the error includes statistical
and all systematic errors added in quadrature.
These values contain the previous estimates obtained with the dynamical
$u$ and $d$ quarks within the error. 
\end{abstract}

\pacs{Valid PACS appear here}% PACS, the Physics and Astronomy
                             % Classification Scheme.

\maketitle

%%%%%%%%%%%%%%%%%%%%

The masses of light quarks are fundamental parameters of QCD.
They cannot be measured experimentally since quarks are confined in hadrons. 
Lattice QCD enables calculations of hadron masses as
functions of quark masses, and hence allows a determination of the quark
masses from the experimental hadron masses.
This approach has been successfully applied, 
first in quenched QCD~\cite{Aoki:1999yr}
and then in $N_f=2$ QCD where degenerate up ($u$) and down ($d$) quarks
are treated dynamically~\cite{AliKhan:2001tx}.  
These studies have revealed that the light quark mass values are 
significantly reduced by dynamical $u$ and $d$ quark effects. 
In this article, we present our attempt to determine the quark masses in
$N_f=2+1$ QCD where the heavier strange ($s$) quark is also treated
dynamically. 
We wish to examine to what extent the dynamical $s$ quark affects
the light quark masses.
We determine the quark masses in the
continuum limit and estimate all possible systematic errors.  
We also calculate the prerequisite light meson spectrum.  
A similar attempt has been made
by the MILC Collaboration~\cite{Bernard:2006wx}.

We adopt the Iwasaki RG gauge action~\cite{Iwasaki:1985we}
and the clover quark action with the improvement coefficient $c_{SW}$
determined nonperturbatively for the RG action~\cite{Aoki:2005et}.
The choice of the gauge action is made to avoid a first-order phase
transition (lattice artifact) observed for the plaquette gauge 
action \cite{Aoki:2004iq}.
We employed the Wilson quark formalism because we prefer an unambiguous
quark-flavor interpretation over the computational ease of the
staggered formalism adopted by the MILC
collaboration~\cite{Aubin:2004ck}. 

\begin{table*}
\caption{\label{TAB:simulation_parameters}
Simulation parameters; $L^3\times T$ is the lattice size,
$(\kappa_{ud},\kappa_s)$ is the hopping parameter combination, 
$1/\delta\tau$ is the number of molecular dynamics steps in one
trajectory, $N_{\rm poly}$ is the PHMC polynomial order, and  
traj. is analyzed trajectory length. Pseudoscalar vector mass ratios
$\frac{m_{\rm PS}}{m_{\rm V}}$ are also listed for light-light (LL)
and strange-strange (SS) mass combinations.
}
\begin{ruledtabular}
\begin{tabular}{ccccccc|ccccccc}
\multicolumn{14}{c}{$\beta=1.83,\,L^3\times T=16^3\times32,\,c_{SW}=1.761$}\\
\hline
$\kappa_{ud}$ & $\kappa_s$ & $\delta\tau$ & $N_{\rm poly}$ & traj. &
$\frac{m_{\rm PS}}{m_{\rm V}}$(LL) & $\frac{m_{\rm PS}}{m_{\rm V}}$(SS) &
$\kappa_{ud}$ & $\kappa_s$ & $\delta\tau$ & $N_{\rm poly}$ & traj. &
$\frac{m_{\rm PS}}{m_{\rm V}}$(LL) & $\frac{m_{\rm PS}}{m_{\rm V}}$(SS)\\ \hline
0.13655 & 0.13710 & 1/80  & 80  & 7000 & 0.7772(13) & 0.7522(15) &
0.13655 & 0.13760 & 1/90  & 110 & 7000 & 0.7769(14) & 0.7235(19)\\
0.13710 &         & 1/85  & 80  & 7000 & 0.7524(21) & 0.7524(21) &
0.13710 &         & 1/100 & 110 & 8600 & 0.7448(14) & 0.7128(16)\\
0.13760 &         & 1/100 & 100 & 7000 & 0.7076(18) & 0.7414(17) &
0.13760 &         & 1/110 & 120 & 8000 & 0.7033(18) & 0.7033(18)\\
0.13800 &         & 1/120 & 110 & 8000 & 0.6629(22) & 0.7365(16) &
0.13800 &         & 1/120 & 130 & 8100 & 0.6525(23) & 0.6941(20)\\
0.13825 &         & 1/140 & 120 & 8000 & 0.6213(24) & 0.7343(15) &
0.13825 &         & 1/150 & 150 & 8100 & 0.6083(32) & 0.6884(21)\\ \hline\hline
\multicolumn{14}{c}{$\beta=1.90,\,L^3\times T=20^3\times40,\,c_{SW}=1.715$}\\
\hline
$\kappa_{ud}$ & $\kappa_s$ & $\delta\tau$ & $N_{\rm poly}$ & traj. &
$\frac{m_{\rm PS}}{m_{\rm V}}$(LL) & $\frac{m_{\rm PS}}{m_{\rm V}}$(SS) &
$\kappa_{ud}$ & $\kappa_s$ & $\delta\tau$ & $N_{\rm poly}$ & traj. &
$\frac{m_{\rm PS}}{m_{\rm V}}$(LL) & $\frac{m_{\rm PS}}{m_{\rm V}}$(SS)\\ \hline
0.13580 & 0.13580 & 1/125 & 110 & 5000 & 0.7673(15) & 0.7673(15) &
0.13580 & 0.13640 & 1/125 & 140 & 5200 & 0.7667(16) & 0.7211(21)\\
0.13610 &         & 1/125 & 110 & 6000 & 0.7435(18) & 0.7647(17) &
0.13610 &         & 1/125 & 140 & 8000 & 0.7444(15) & 0.7182(17)\\
0.13640 &         & 1/140 & 110 & 7600 & 0.7204(19) & 0.7687(15) &
0.13640 &         & 1/140 & 140 & 9000 & 0.7145(16) & 0.7145(16)\\
0.13680 &         & 1/160 & 110 & 8000 & 0.6701(27) & 0.7673(17) &
0.13680 &         & 1/160 & 140 & 9200 & 0.6630(21) & 0.7127(17)\\
0.13700 &         & 1/180 & 110 & 7900 & 0.6390(22) & 0.7691(15) &
0.13700 &         & 1/180 & 140 & 7900 & 0.6243(28) & 0.7102(20)\\ \hline\hline
\multicolumn{14}{c}{$\beta=2.05,\,L^3\times T=28^3\times56,\,c_{SW}=1.628$}\\
\hline
$\kappa_{ud}$ & $\kappa_s$ & $\delta\tau$ & $N_{\rm poly}$ & traj. &
$\frac{m_{\rm PS}}{m_{\rm V}}$(LL) & $\frac{m_{\rm PS}}{m_{\rm V}}$(SS) &
$\kappa_{ud}$ & $\kappa_s$ & $\delta\tau$ & $N_{\rm poly}$ & traj. &
$\frac{m_{\rm PS}}{m_{\rm V}}$(LL) & $\frac{m_{\rm PS}}{m_{\rm V}}$(SS)\\ \hline
0.13470 & 0.13510 & 1/175 & 200 & 6000 & 0.7757(26) & 0.7273(29) & 
0.13470 & 0.13540 & 1/175 & 250 & 6000 & 0.7790(23) & 0.6821(32)\\
0.13510 &         & 1/195 & 200 & 6000 & 0.7316(24) & 0.7316(24) &
0.13510 &         & 1/195 & 250 & 6000 & 0.7341(29) & 0.6820(39)\\
0.13540 &         & 1/225 & 200 & 6000 & 0.6874(30) & 0.7395(23) &
0.13540 &         & 1/225 & 250 & 6000 & 0.6899(34) & 0.6899(34)\\
0.13550 &         & 1/235 & 200 & 6500 & 0.6611(34) & 0.7361(25) &
0.13550 &         & 1/235 & 250 & 6500 & 0.6679(45) & 0.6899(43)\\
0.13560 &         & 1/250 & 200 & 6500 & 0.6337(38) & 0.7377(28) &
0.13560 &         & 1/250 & 250 & 6500 & 0.6361(47) & 0.6852(46)\\
\end{tabular}
\end{ruledtabular}
\vspace*{-2mm}
\end{table*}

Configurations are generated at three values of the coupling
$\beta\equiv6/g^2=2.05, 1.90$ and $1.83$ corresponding to the 
squared lattice spacing $a^2\simeq0.005, 0.01$ and $0.015$ ${\rm fm}^2$,
with the physical volume fixed to about $(2.0{\rm fm})^3$.
At each $\beta$, we perform simulations for $10$ quark mass combinations
using a combined algorithm~\cite{Aoki:2001pt} of the Hybrid Monte
Carlo (HMC) for the degenerate $u$ and $d$ quarks and 
the polynomial Hybrid Monte Calro
(PHMC) for the $s$ quark. 
Table \ref{TAB:simulation_parameters} summarizes the simulation parameters. 

The meson and quark masses at the simulation points are determined from
single exponential correlated $\chi^2$ fits to the correlators 
$\langle P(t)P(0)\rangle$, $\langle V(t)V(0)\rangle$ and  
$\langle A_4(t)P(0)\rangle$, where $P$, $V$ and $A_{\mu}$ denote
pseudoscalar, vector and nonperturbatively 
$O(a)$-improved~\cite{Kaneko:2007wh} axial-vector current operators, 
respectively. 
We use an exponentially smeared source and a point sink, and measurements 
are made at every $10$ HMC trajectories in the Coulomb gauge. 
For the pseudoscalar sector, $\langle P(t)P(0)\rangle$ and 
$\langle A_4(t)P(0)\rangle$ are fitted simultaneously ignoring 
correlations among them.  
Errors are estimated by the jackknife method with a bin size of
$100$ HMC trajectories; errors do not increase for larger bin sizes.

\begin{figure}[t]
\begin{center}
\hspace*{-5mm}
\parbox{40mm}{
\includegraphics[scale=0.35, viewport = 0 0 330 510, clip]
{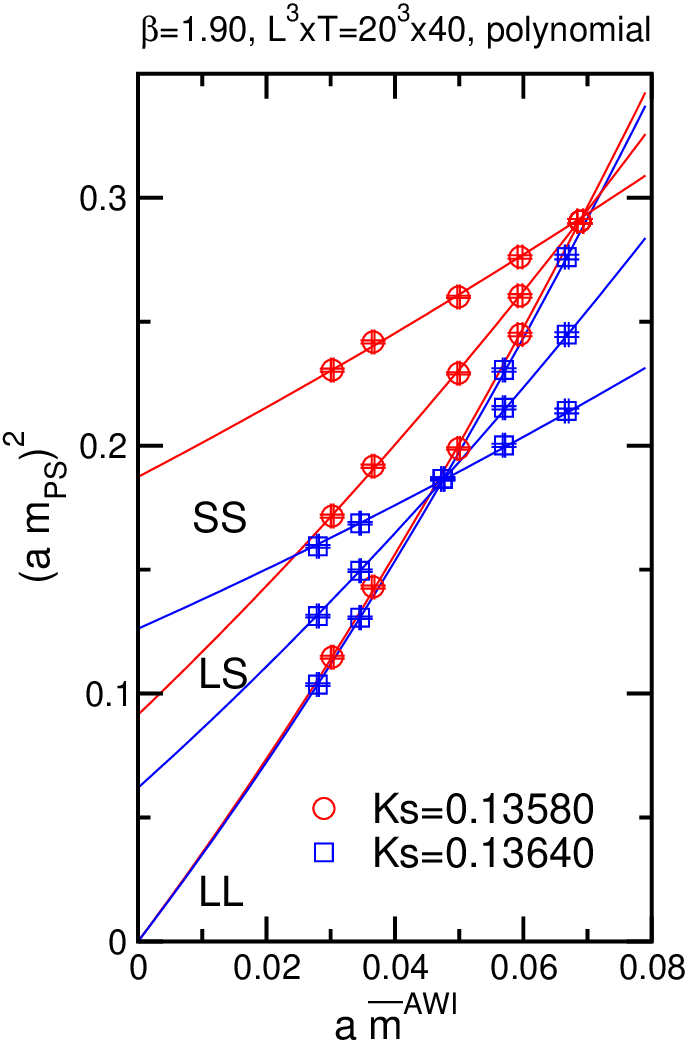}
}
\hspace*{+2mm}
\parbox{40mm}{
\includegraphics[scale=0.35, viewport = 0 0 330 510, clip]
{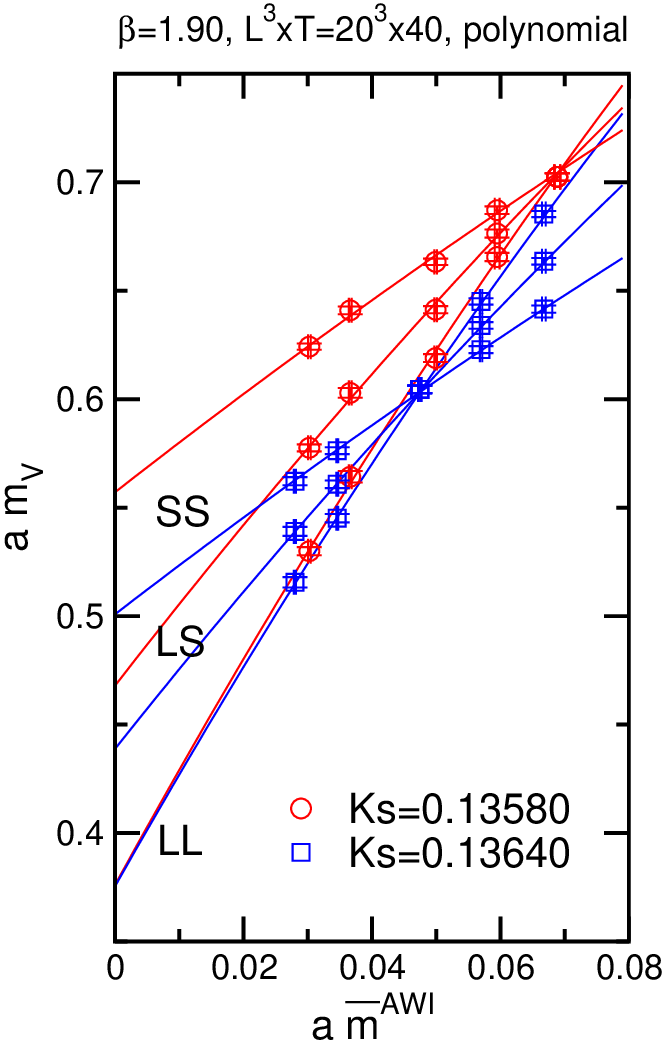}
}
\vspace*{-1mm}
\caption{
Chiral fits of meson masses with $m_q^{AWI}$ at $\beta=1.90$.
}
\label{FIG:Chiral_fit_Poly}
\end{center}
\vspace*{-10mm}
\end{figure}

Chiral fits are made to the light-light (LL), light-strange (LS) and
strange-strange (SS) meson masses simultaneously ignoring their
correlations, using a quadratic polynomial function of
the sea quark masses $(m_u, m_d, m_s)$ and valence quark masses
$(m_{{\rm val} 1}, m_{{\rm val} 2})$ in mesons;
\begin{eqnarray}
&&f({\rm M_s}, {\rm M_v}) \label{eq:chiralfit}\\
&=&A+B_S{\rm tr}{\rm M_s}+B_V{\rm tr}{\rm M_v}
+D_{SV}{\rm tr}{\rm M_s}{\rm tr}{\rm M_v}\nonumber\\
&&+C_{S1}{\rm tr}{\rm M_s^2}+C_{S2}({\rm tr}{\rm M_s})^2
+C_{V1}{\rm tr}{\rm M_v^2}+C_{V2}({\rm tr}{\rm M_v})^2,\nonumber
\end{eqnarray}
where $f=m_{PS}^2$ or vector meson mass $m_V$, 
${\rm M_S}={\rm diag}(m_u, m_d, m_s)$,
${\rm M_V}={\rm diag}(m_{{\rm val} 1}, m_{{\rm val} 2})$,
and ``${\rm tr}$'' means the trace of matrices.
In the fits, we use the axial-vector Ward identity
quark mass $m_q=\lim_{t\rightarrow\infty}
\langle\partial_4A_4(t)P(0)\rangle/(2\langle P(t)P(0)\rangle)$
and set $A=B_S=C_{S1}=C_{S2}=0$ for $m_{PS}^2$.
These fits reproduce measured data well, as illustrated in
Fig.\ref{FIG:Chiral_fit_Poly},
with reasonable $\chi^2/{\rm d.o.f.}$ of at most $1.36$.

%%%%%%%%%%%%%%%%%%%%
%%%%%%%%%%%%%%%%%%%%

\begin{figure}
\includegraphics[scale=0.42, viewport = 0 0 520 510, clip]
{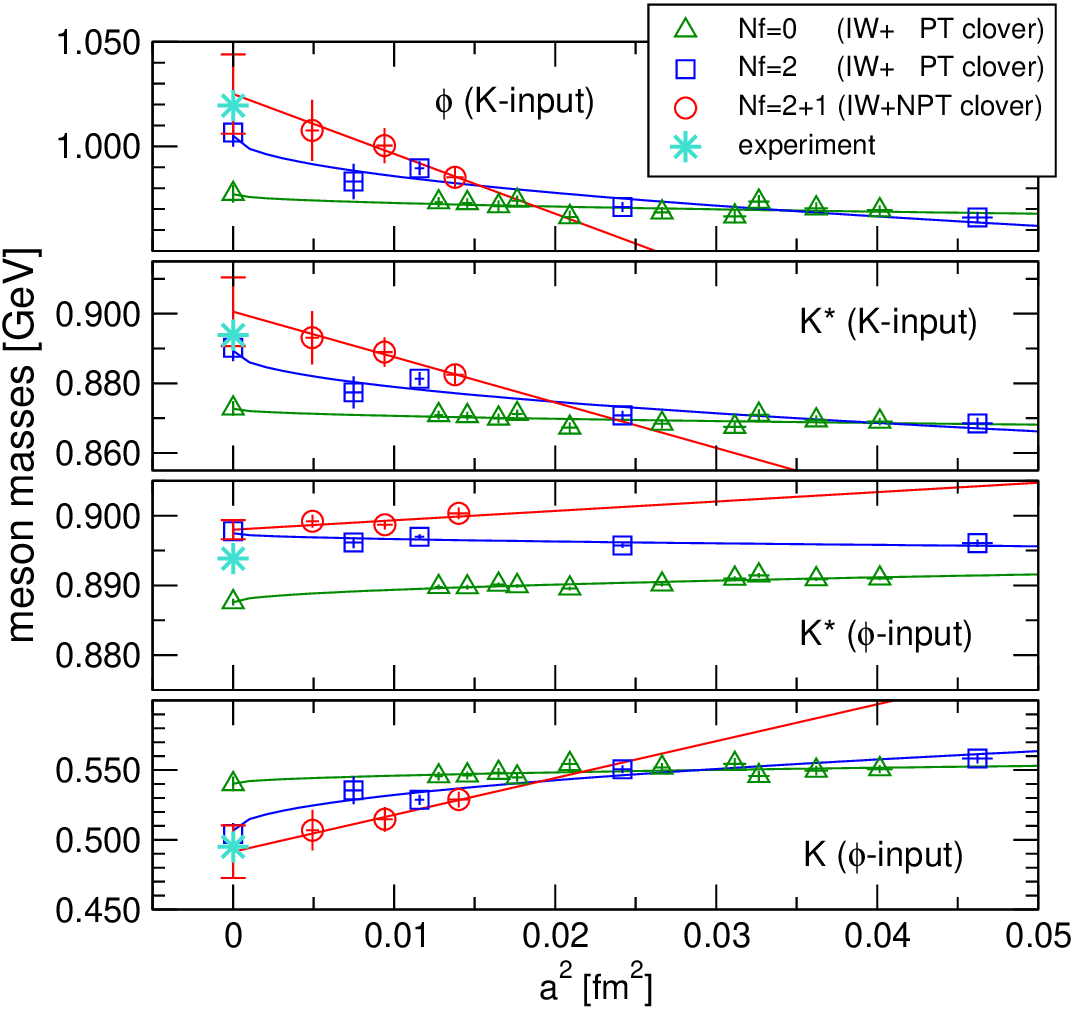}
\vspace*{-3mm}
\caption{\label{FIG:meson_masses}
Continuum extrapolation of meson masses % (from VWI quark mass fits) 
for $N_f=2+1$ QCD (circles), compared to experiment (stars)
and results in $N_f=2$ (squares) and $N_f=0$ (triangles) 
QCD~\cite{AliKhan:2001tx}.
}
\vspace*{-3mm}
\end{figure}

The physical quark mass point and the lattice spacing are determined from
the experimental values of $\pi^0$, $\rho^0$ and $K$ ($K$-input) or
$\pi^0$, $\rho^0$ and $\phi$ ($\phi$-input) meson masses.
Taking the $\rho^0$ mass as input may cause a large
systematic error, because the $\rho\to\pi\pi$ decay mode is
not open for our mass range
(the lightest pion mass in this simulation $\sim 620$~MeV)
and hence chiral extrapolation
of $m_V$ for lighter quarks may be quite different from our fits. 
In order to estimate this uncertainty, we also
check another combination $[\pi^0, K, \phi]$.
We assume the ideal mixing for the vector isosinglets. 
Since our simulation is made with degenerate $u$ and $d$ quarks, we
consider the isospin averages
$m_{\hat{K}}=\{(m_{K^{\pm}}^2+m_{K^0}^2)/2\}^{1/2}$ and
$m_{\hat{K}^{\ast}}=(m_{K^{\ast\pm}}+m_{K^{\ast 0}})/2$
and predict the average light quark mass $\overline{m}=(m_u+m_d)/2$. 
The electromagnetic (EM) effects, not included in our
simulations, are removed from the $m_{K^{\pm}}$ above using
Dashen's theorem~\cite{Dashen:1969eg}
$(m_{K^{\pm}}^2-m_{K^0}^2)_{\rm EM}
=(m_{\pi^{\pm}}^2-m_{\pi^0}^2)_{\rm EXP}$.
The isospin breaking effects and the EM effects for other
mesons we consider are expected to be small and thus are not
considered.  
The experimental values we use are taken
from the PDG booklet~\cite{Yao:2006px};
$m_{\pi^0}=0.1350 \mbox{GeV}$,
$m_{\pi^{\pm}}=0.1396 \mbox{GeV}$,
$m_{K^0}=0.4976 \mbox{GeV}$,
$m_{K^{\pm}}=0.4937 \mbox{GeV}$,
$m_{\rho^0}=0.7755 \mbox{GeV}$,
$m_{K^{\ast 0}}=0.8960 \mbox{GeV}$,
$m_{K^{\ast\pm}}=0.8917 \mbox{GeV}$ and
$m_{\phi}=1.0195 \mbox{GeV}$. 
Lattice spacings (Table~\ref{TAB:lattice_spacings})
for the $K$- and $\phi$- inputs are consistent,
while those for the $[\pi, K, \phi]$-input 
are slightly smaller by at most 7\%.\\
\begin{table}
\caption{
\label{TAB:lattice_spacings}
Lattice spacings in fm units.
}
\begin{ruledtabular}
\begin{tabular}{c|cc|c}
$\beta$ & $K$-input & $\phi$-input & $[\pi,K,\phi]$-input\\ \hline
1.83 & 0.1174(23) & 0.1184(26) & 0.1095(25)\\
1.90 & 0.0970(26) & 0.0971(25) & 0.0936(33)\\
2.05 & 0.0701(29) & 0.0702(28) & 0.0684(41)\\
\end{tabular}
\end{ruledtabular}
\vspace*{-3mm}
\end{table}
\begin{table}
\caption{
\label{TAB:meson_spectrum}
Meson masses in the continuum limit (in MeV units),
compared to experiment.
The EM effect is subtracted using Dashen's theorem.
}
\begin{ruledtabular}
\begin{tabular}{l|cc|c||c}
& $K$-input & $\phi$-input &
$[\pi,K,\phi]$-input & EXP. \\
\hline
$\hat K$         & -          & 491(19)    & -       & 495.0 \\
$\rho^0$         & -          & -          & 761(32) & 775.5 \\
$\hat K^{\ast}$  & 900.5(9.9) & 898.0(1.4) & 891(16) & 893.9 \\
$\phi$           & 1025(19)   & -          & -       & 1019.5\\
\end{tabular}
\end{ruledtabular}
\vspace*{-3mm}
\end{table}
An agreement of the meson spectrum with experiment is a necessary condition
for a reliable estimate of the quark masses.
To confirm this, we extrapolate the meson masses linearly in $a^2$,
because our action is $O(a)$ improved and data are well fitted, as
shown in Fig.\ref{FIG:meson_masses}, with small 
$\chi^2/{\rm d.o.f}\le 1.4$.
The masses in the continuum limit, summarized
in Table \ref{TAB:meson_spectrum}, 
are consistent with experiment with at most $2.9\sigma$ deviation.
The $\hat{K}^{\ast}$ mass turns out to be slightly heavier than
experiment, though the supplemental $[\pi,K,\phi]$-input
gives consistent results with experiment with large statistical error.
Possible origin of the deviation is due to uncertainty of
chiral fits. In fact, an alternative fit based on chiral
perturbation theory ($\chi$PT) we discuss later yields
$m_{\hat{K}^{\ast}}=894(12)$ MeV ($K$-input). 
In Fig.\ref{FIG:meson_masses} we overlay the previous results of meson
masses~\cite{AliKhan:2001tx} in the $N_f=2$ and quenched ($N_f=0$) QCD
with tadpole improved one-loop $c_{SW}$.
The dynamical $u$ and $d$ quarks significantly reduce the $O(10\%)$
deviation of the quenched spectrum from experiment. 
We find no further dynamical $s$ quark effect beyond statistical errors.

%%%%%%%%%%%%%%%%%%%%
%%%%%%%%%%%%%%%%%%%%

%%%%%%%%%%%%%%%%%%%%
\begin{figure}[t]
\begin{center}
\parbox{40mm}{
\includegraphics[scale=0.37, viewport = 0 0 300 480, clip]
{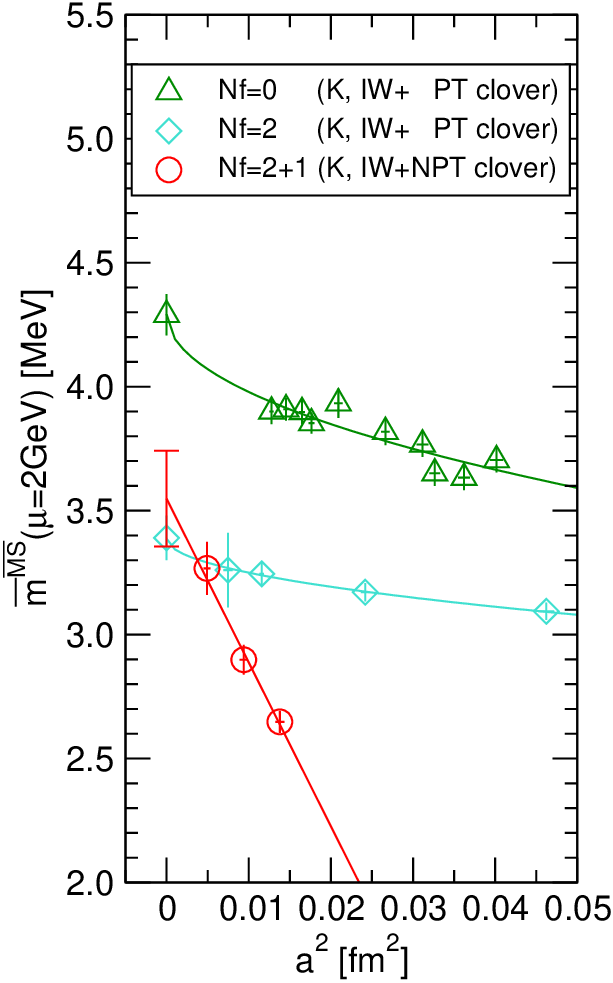}
}
\hspace*{+2mm}
\parbox{40mm}{
\includegraphics[scale=0.37, viewport = 0 0 300 480, clip]
{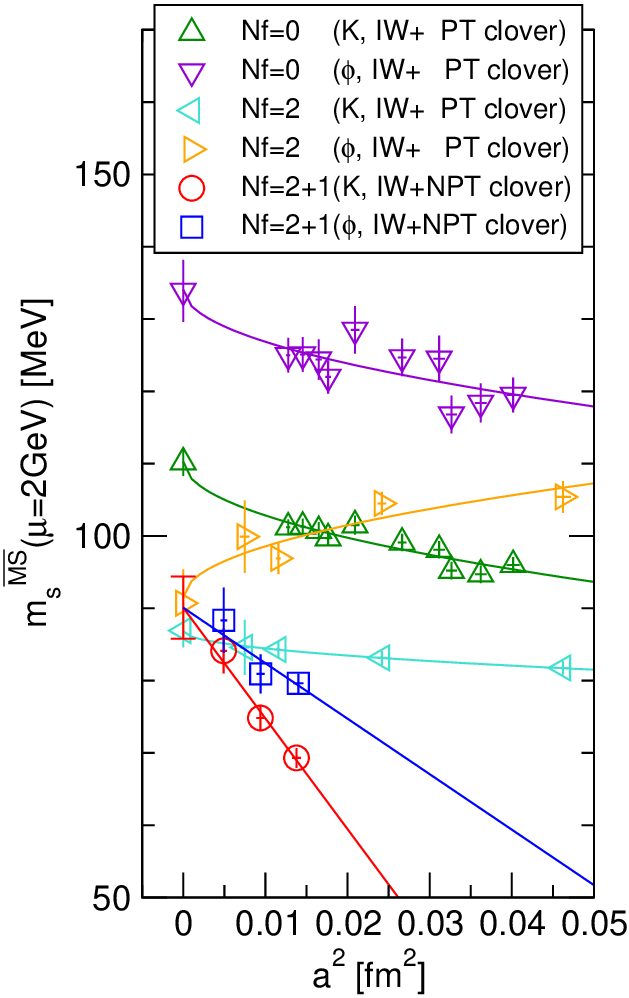}
}
\vspace*{-1mm}
\caption{Continuum extrapolations of the up, down and strange quark masses.
         For comparison, results for $N_f=0$ and $N_f=2$ QCD
         ~\protect\cite{AliKhan:2001tx} are overlaid.
         }
\label{FIG:uds_quark_masses}
\end{center}
\vspace*{-8mm}
\end{figure}
%%%%%%%%%%%%%%%%%%%%

The quark masses are evaluated for the $\overline{\mbox{MS}}$ scheme at
the scale $\mu=2 {\rm GeV}$ using the tadpole improved one-loop 
matching~\cite{Aoki:1998ar} at $\mu=a^{-1}$ with an improved coupling 
determined from plaquette and rectangular loop and four-loop
renormalization group equation.
In the continuum extrapolation of the quark masses, we assume
the $O(g^4ma)$ contributions are small and neglect it.
As Fig.~\ref{FIG:uds_quark_masses} shows, the quark masses are well
described by a linear function in $a^2$, and the values determined for
either the $K$- or the $\phi$-inputs, while different at finite lattice
spacings, extrapolate to a common value in the continuum limit.  
Therefore the continuum limit is estimated from 
a combined linear fit with the $K$- and the $\phi$-inputs.
We obtain 
$\overline{m}^{\overline{\rm MS}}(\mu=2{\rm GeV})=3.55(19)~{\rm MeV}$ and
$m_s^{\overline{\rm MS}}(\mu=2{\rm GeV})=90.1(4.3)~{\rm MeV}$ 
with a sufficiently small $\chi^2/{\rm d.o.f.}< 0.42$.
Note that the supplemental $[\pi,K,\phi]$-input gives larger
statistical error and hence is not used to estimate central values.

We now turn to estimates of possible systematic errors. \\
{\bf Finite size effect (FSE)} ---
The meson masses at the infinite volume are estimated at $\beta=1.90$ 
using data on a $V\sim(2.0{\rm fm})^3$ lattice and those from
our exploratory study on a $V\sim(1.6{\rm fm})^3$ 
lattice~\cite{Kaneko:2003re},
and assuming a strong volume dependence of  
$(m_{{\rm had},V}-m_{{\rm had},V=\infty})/m_{{\rm had},V=\infty}\propto
1/V$~\cite{Fukugita:1992jj}.
The chiral fits to the infinite volume values lead to less than a $4\%$
change for the meson masses at the physical point. 
For the quark masses, however, we find a larger shift of $12.2\%$ from
a $V\sim(2.0{\rm fm})^3$ lattice to $V=\infty$ for
$\overline{m}$ with $\phi$-input and $8.1\%$ for $m_s$ with $K$-input
(differences are smaller for the other cases). 
Assuming that FSE is independent of lattice spacing, we take the
differences as estimates of FSE for the quark masses in the continuum limit.\\
{\bf Chiral extrapolation} ---
In addition to the polynomial chiral fits, we fit the meson
masses using $\chi$PT formulaes modified for the Wilson quark action
(W$\chi$PT)~\cite{Sharpe:1998xm}.
Namely, we fit $m_{\pi}$, $m_{\hat{K}}$, $m_{\rho}$ and $m_{\hat{K}^{\ast}}$ 
using the NLO $N_f=2+1$ QCD W$\chi$PT
formulae for the $O(a)$ improved theory~\cite{Aoki:2005mb}.  
Since the formula in Ref.~\cite{Aoki:2005mb} is not applicable
for the $\phi$ meson, we estimate the effect only for $K$-input.
In the fits we obtain $\overline{m}$ to be $3.1\%$ smaller and $m_s$
to be $1.2\%$ larger than those of the polynomial fit. 
We note that our W$\chi$PT fits to data do not exhibit a clear chiral
logarithm, probably because $u$ and $d$ quark masses
in our simulation are not sufficiently small. 
Further possible systematic error from a long
chiral extrapolation for $\rho^0$, mentioned above, is
estimated by the supplemental $[\pi,K,\phi]$-input, which
gives $3.0\%$ larger for $\overline{m}$ and
$3.4\%$ larger value for $m_s$ than the central one.
For an estimate of systematic errors from chiral fits, 
differences of the two alternative fits from the central
value are added linearly. \\
{\bf Renormalization factor} ---
Uncertainty of the one-loop calculation of the renormalization factor
is estimated by shifting the matching scale from $\mu=1/a$ to 
$\mu=\pi/a$ and also using an alternative tadpole improved coupling
~\cite{AliKhan:2001tx}.\\
{\bf Continuum extrapolation} ---
Possible $O(a^3)$ effects are investigated by performing the continuum
extrapolation adding an $O(a^3)$ term to the fit function.\\
{\bf Electromagnetic (EM) effects} ---
Systematic error due to uncertainty of the EM effects is
estimated following extensive arguments
~\cite{Bijnens:1996kk, Aubin:2004ck, Nelson:2002si} to Dashen's
theorem~\cite{Dashen:1969eg}.
Namely, we estimate the effects by
a further mass shift of our input $m_{\hat{K}}$
using a relation $(m_{K^{\pm}}^2-m_{K^0}^2)_{\rm EM}
=(1+\Delta_E)(m_{\pi^{\pm}}^2-m_{\pi^0}^2)_{\rm EXP}$
assuming the EM effects for other mesons are negligeble. 
We vary the $\Delta_E$ in range $[-1, +1]$
as our estimate of the EM effects,
and we find a quite small change in $m_s$
and no change in $\overline{m}$.\\
{\bf Isospin breaking effects} ---
Isospin breaking effects are estimated by chiral fits with
Eq.~(\ref{eq:chiralfit}) for $m_u\ne m_d$ and taking 
$m_{\pi^0}$, $m_{\rho^0}$, $m_{K^\pm}$ and $m_{K^0}$ as inputs.
We find that $m_u/m_d = 0.577(25)$,
and that $\overline{m}$ and $m_s$ have no change from
the $\hat{K}$ input result.
We note that $m_u/m_d$ strongly depends on an
estimate of the EM effects; $m_u/m_d=$$0.663$--$0.498$ for
$\Delta_E=[-1, +1]$, though $\overline{m}$ and $m_s$ almost do not. 

Finally we obtain
\begin{eqnarray}
&&\overline{m}^{\overline{\rm MS}}(\mu=2{\rm GeV})\nonumber\\
&&=3.55(19)({}^{+43}_{-0})({}^{+11}_{-11})({}^{+26}_{-17})
({}^{+34}_{-0})({}^{+0}_{-0})({}^{+0}_{-0}),\\
&&m_s^{\overline{\rm MS}}(\mu=2{\rm GeV})\nonumber\\
&&=90.1(4.3)({}^{+7.3}_{-0})({}^{+4.2}_{-0})({}^{+6.6}_{-4.3})
({}^{+12.8}_{-0})({}^{+0.1}_{-0.2})({}^{+0}_{-0}),
\end{eqnarray}
in MeV units, where the errors are statistical, systematic due to FSE,
chiral extrapolation, renormalization factor, continuum extrapolation,
EM effect and isospin breaking effect, respectively.
Adding the errors in quadrature yields the values quoted in the abstract.
These values agree well with the latest report from the MILC
Collaboration~\cite{Bernard:2006wx}
$\overline{m}=3.3\pm 0.3$~MeV and $m_s=90\pm 6$~MeV where we added 
the quoted errors in quadrature.
They also include the $N_f=2$ values~\cite{AliKhan:2001tx}
within the error.

Scaling violation in the quark masses is unexpectedly large, while 
that for the meson masses are reasonably bounded at a percent level 
at $a\approx 0.1~{\rm fm}$.
To gain a better control over systematic uncertainties,
a significant reduction in the simulated light quark masses 
on a correspondingly larger lattice is needed.  
An attempt is underway to meet these challenges
~\cite{Kuramashi:2006np}.\\

\begin{acknowledgments}
This work is supported by 
the Epoch Making Simulation Projects of Earth Simulator Center,
the Large Scale Simulation Program No.132 (FY2005) of 
High Energy Accelerator Research Organization (KEK),
the Large Scale Simulation Projects of 
Academic Computing and Communications Center of University of Tsukuba,
Inter University Services of Super Computers of 
Information Technology Center of University of Tokyo,
Super Sinet Projects of National Institute of Informatics,
and also by the Grant-in-Aid of the Ministry of Education
(Nos. 13135204, 13135216, 15540251, 16540228, 16470147, 17340066, 17540259, 
18104005, 18540250, 18740130).
\end{acknowledgments}

%%%%%%%%%%%%%%%%%%%%%%%%%%%%%%%%%%%%%%%%%%%%%%%%%%%%%%%%%%%
%%%%%%%%%%%%%%%%%%%%%% References %%%%%%%%%%%%%%%%%%%%%%%%%
%%%%%%%%%%%%%%%%%%%%%%%%%%%%%%%%%%%%%%%%%%%%%%%%%%%%%%%%%%%

\end{document}